\documentclass[aps,showpacs,nofootinbib,superscriptaddress]{revtex4-1}
\usepackage{epsf}
\usepackage{graphicx}
\usepackage{amsmath}
\def\slashchar#1{\setbox0=\hbox{$#1$}
   \dimen0=\wd0 \setbox1=\hbox{/} \dimen1=\wd1
   \ifdim\dimen0>\dimen1 \rlap{\hbox to \dimen0{\hfil/\hfil}} #1
   \else  \rlap{\hbox to \dimen1{\hfil$#1$\hfil}} / \fi}

\newcommand{\be}{\begin{equation}}
\newcommand{\ee}{\end{equation}}
\newcommand{\bea}{\begin{eqnarray}}
\newcommand{\eea}{\end{eqnarray}}

\begin{document}

\title{ Hyperfine mixing in electromagnetic decay of doubly
heavy $bc$ baryons }

\author{ C.Albertus} \affiliation{Departamento de F\'\i
sica Fundamental, Universidad de Salamanca, E-37008 Salamanca,
Spain} \author{
E. Hern\'andez} \affiliation{Departamento de F\'\i sica Fundamental e
IUFFyM,\\ Universidad de Salamanca, E-37008 Salamanca, Spain}
\author{J.~Nieves} \affiliation{Instituto de F\'\i sica Corpuscular
(IFIC), Centro Mixto CSIC-Universidad de Valencia, Institutos de
Investigaci\'on de Paterna, Aptd. 22085, E-46071 Valencia, Spain}


\begin{abstract}
  We investigate the role of hyperfine mixing in the electromagnetic
  decay of ground state doubly heavy $bc$ baryons. As in the case of a
  previous calculation on $b\to c$ semileptonic decays of doubly heavy
  baryons, we find large corrections to the electromagnetic decay
  widths due to this mixing. Contrary to the weak case just mentioned, we
  find here that one can not use electromagnetic width relations
  obtained in the infinite heavy quark mass limit to experimentally
  extract information on the admixtures in a model independent way.
\end{abstract}
\maketitle
%
%
%
%
%
%
\section{Introduction}
In the infinite heavy quark mass limit, and according to heavy quark
 spin symmetry (HQSS)~\cite{Jenkins:1992nb}, one can select the heavy
 quark subsystem of a doubly heavy baryon to have a well defined total
 spin $S_h=0,1$. This has been the default assumption by most
 calculations of doubly heavy baryon
 spectroscopy~\cite{korner94,silvestre96,ebert97,itoh00,gershtein00,tong00,
 mathur02,
 ebert02,kiselev02,narodetskii02,vijande04,albertusdhb,zhang2008}. However, due
 to the finite value of the heavy quark masses, the hyperfine
 interaction between the light quark and any of the heavy quarks can
 admix both $S_h=0$ and $S_h=1$ spin components into the wave
 function.  For ground state (total orbital angular momentum $L=0$)
 $bc$ baryons, one should expect the actual physical $\Xi$ particles
 (quark content $bcu$ or $bcd)$ to be mixtures of the
 $\Xi_{bc}\,(S_h=1)$ and $\Xi'_{bc}\, (S_h=0)$ states. Similarly in
 the strange sector, the physical $\Omega$ particles (quark content
 $bcs$) will be mixtures of $\Omega_{bc}\,(S_h=1)$ and
 $\Omega'_{bc}\,(S_h=0)$) states.

While mixing effects are negligible in the spectrum, it was pointed
out in Ref.~\cite{pervin1} that hyperfine mixing could greatly affect
the decay widths of doubly heavy baryons. The calculation for $b\to c$
semileptonic decay of doubly heavy baryons was conducted by the same
authors in Ref.~\cite{pervin2}, where they found that hyperfine mixing
in the $bc$ states had a tremendous impact on the decay widths.  We
qualitatively confirmed their results in Ref.~\cite{albertus2010},
although our predictions for the decay widths were roughly a factor of
two larger.  There, we also showed how HQSS predictions for $b\to c$
semileptonic decay, could be used to experimentally obtain information
on the admixtures of the $bc$ baryons in a model independent
manner. Unfortunately those ratios involved weak decays that have
competing electromagnetic (e.m.) decays and thus, they will be
difficult to observe experimentally. In this context, it was clear
 the possible relevance of hyperfine mixing effects in
e.m. decays.  In fact, the authors of Ref.~\cite{pervin2} expected
hyperfine mixing effects to play an important role also for
e.m. transitions. As a result of these considerations we included in
Ref.~\cite{albertus2010} predictions for ratios that involved
e.m. decay widths evaluated in the infinite heavy quark mass
limit. This limit implies that the spin of the heavy quark subsystem
can not change in an e.m. transition.
 
 In this letter we perform the full calculation using the same quark
model as in Ref.~\cite{albertus2010}. To our knowledge there is only one prior
 calculation 
of e.m.
decays of doubly heavy $bc$ baryons~\cite{Dai:2000hza}. There, the authors used the
e.m. radiation as a means to investigate the diquark structure
but  no hyperfine mixing was  considered.
In this work we restrict ourselves to
transitions involving ground state ($L=0$) $bc$ baryons and  our emphasis 
 is put on  the relevance of hyperfine
 mixing for
those transitions. In
Table~\ref{tab:dhb} we show the quantum numbers for the ground state
of unmixed $bc$ baryons classified so that $S_h$ is well defined. The
physical spin-1/2 $bc$ states are mixtures of the
$\Xi_{bc},\,\Xi_{bc}'$ ($\Omega_{bc},\,\Omega_{bc}'$) states shown in that
table.  Their quantum numbers and admixture coefficients appear in
Table~\ref{tab:dhbphys}.   As for the weak $b\to c$ decays analyzed in
Ref.~\cite{albertus2010}, we find here that hyperfine mixing largely
affects the e.m. decay widths. On the other hand we find that
contributions that change the spin of the heavy quark subsystem are
very important in the evaluation of the e.m. decay widths. We are thus
far from the infinite heavy quark mass limit according to which the
spin of the heavy quark subsystem can not change in an
e.m. transition. Due to this fact the e.m. decay width ratios proposed
in Ref.~\cite{albertus2010}, and obtained within that assumption, are
not valid for the actual heavy quark masses and can not be used to
experimentally extract information on the admixtures in a model
independent way.

The paper is organized as follows: in Sec.~\ref{sec:emdecay} we
 collect general formulas to evaluate the e.m. decay width. We also
 give an appropriate form factor decomposition of the electromagnetic
 current matrix elements as well as showing how the form factors can
 be obtained in terms of those matrix elements. In
 Sec.~\ref{sec:model} we present our nonrelativistic states and the
 way the matrix elements are evaluated in our model.  Finally in
 Sec.~\ref{sec:results} we present the results and the conclusions of
 our work.

 \begin{table}
\begin{tabular}{ccccc||ccccc}\hline
\hspace{.25cm}Baryon\hspace{.25cm} & Quark content & \hspace{.25cm}$S_h^\pi$\hspace{.25cm} & 
\hspace{.25cm}$J^\pi$\hspace{.25cm} & M\ [{\rm MeV}]
&\hspace{.25cm}Baryon\hspace{.25cm} & Quark content & \hspace{.25cm}$S_h^\pi$\hspace{.25cm} &
\hspace{.25cm}$J^\pi$\hspace{.25cm} & M\ [{\rm MeV}]\vspace{-.1cm}\\ 
&(l=u,d)&&&&&&\\ \hline
$\Xi_{bc}^*$ & \{b~c\}~l & 1$^+$ & 3/2$^+$&6996&\hspace{.25cm}$\Omega_{bc}^*$ &
\{b~c\}~s & 1$^+$ & 3/2$^+$& 7075\\ 
$\Xi_{bc}'$ & [b~c]~l & 0$^+$ & 1/2$^+$& 6958&\hspace{.25cm}$\Omega_{bc}'$ & [b~c]~s & 0$^+$
& 1/2$^+$&  7038\\ 
$\Xi_{bc}$ & \{b~c\}~l & 1$^+$ & 1/2$^+$&6928 &\hspace{.25cm}$\Omega_{bc}$ &
\{b~c\}~s & 1$^+$ & 1/2$^+$ & 7013\\ 
\hline
\end{tabular}%
\caption{Quantum numbers and quark content for unmixed ground state
 doubly heavy $bc$ baryons with well defined $S_h$ (spin of the heavy
 quark subsystem). The masses were obtained in
 Ref.~\cite{albertus2010}.  $S_h^\pi$ stands for the spin and parity
 of the heavy quark subsystem, while $J^\pi$ stands for the total spin
 and parity of the baryon. For $J=1/2$, actual physical states are
 mixtures of the $\Xi_{bc},\,\Xi_{bc}'$ ($\Omega_{bc},\,\Omega_{bc}'$)
 states, and they appear in Table~\ref{tab:dhbphys}.}
\label{tab:dhb}
\end{table}
 \begin{table}
\begin{tabular}{ccc||ccc}\hline
\hspace{.25cm}Baryon\hspace{.25cm} &\hspace{.25cm} $J^\pi$\hspace{.25cm} & M\ [{\rm MeV}]
&\hspace{.25cm}Baryon\hspace{.25cm} &\hspace{.25cm} $J^\pi$\hspace{.25cm}  &  M\ [{\rm MeV}]\\ \hline
$\Xi_{bc}^{(1)}=\hspace{.25cm}0.902\ \Xi'_{bc}+0.431\ \Xi_{bc}$ &$1/2^+$ &6967 &
\hspace{.25cm}$\Omega_{bc}^{(1)}=\hspace{.25cm}0.899\  \Omega'_{bc}+0.437\ 
\Omega_{bc}$
&$1/2^+$   & 7046\\ 
$\Xi_{bc}^{(2)}=-0.431\  \Xi'_{bc}+0.902\ \Xi_{bc}$&$1/2^+$  & 6919
&\hspace{.25cm}$\Omega_{bc}^{(2)}=-0.437\  \Omega'_{bc}+0.899\ \Omega_{bc}$&$1/2^+$  &  7005\\ 
\hline
\end{tabular}%
\caption{Physical spin-1/2 doubly heavy $bc$ baryons. The admixture
coefficients and the physical masses were obtained in Ref.~\cite{albertus2010}.
}
\label{tab:dhbphys}
\end{table}
%
%
%
%
%
%
\section{Electromagnetic decay}
\label{sec:emdecay}
The electromagnetic decay width for the $B\to B'\,\gamma$ 
process is given by\footnote{Note the normalization of the baryon states should
be such that
\be
\left\langle B, s' \vec{P}'\,|B, s \vec{P}
\right\rangle=\delta_{s\,s'}\,(2\pi)^3\,2E\,\delta^{(3)}(\vec{P}-\vec{P}')
\ee
}
\begin{eqnarray}
\label{eq:gamma}
\Gamma&=&\frac{1}{2M}\int
\frac{d^{\,3}P'}{(2\pi)^3\,2E'}
\int \frac{d^{\,3}q}{(2\pi)^3\,2\omega}
\ (2\pi)^4\delta^{(4)}(P-P'-q)\nonumber\\
&&\hspace{3cm}\times\ \frac{1}{2 J+1}\sum_s\ \sum_{s'}\sum_r
\left({\cal J}^{BB'\ \mu}_{s\,s'}(P,P')\,\varepsilon_{r\,\mu}(q)\right)
\left({\cal J}^{BB'\ \nu}_{s\,s'}(P,P')\,\varepsilon_{r\,\nu}(q)\right)^*
\end{eqnarray}
where $P=(M;\vec{0}\,)$, $P'=(E'=\sqrt{M'^2+\vec P'^2},\vec{P}')$ are
respectively the four momenta of the initial and final baryons.  $J$
is the total spin of the initial baryon and $s,\,s'$ are the spin
third components of the initial and final baryons. $q=(\omega=|\vec
q\,|,\vec{q}\,)$ is the final photon four momenta, being
$\varepsilon_r(q)$ its polarization vector.  Finally ${\cal J}^{BB'\
\mu}_{s\,s'}(P,P')$ stands for the electromagnetic current matrix
element
\begin{eqnarray}
{\cal J}^{BB'\ \mu}_{s\,s'}(P,P')=
\left\langle B',\ s'\,\vec{P}'\,\left|J_{em}^\mu(0)\,\right|
\,B,\ s\,\vec P=\vec 0\,\right\rangle
\end{eqnarray}
with
\be
J_{em}^\mu(0)=e\sum_q e_q\bar\Psi_q(0)\gamma^\mu\Psi_q(0)\ \ ;\ \
\frac{e^2}{4\pi}=\alpha_{em}
\ee
where the different $e_q$ are the quark charges in units of the proton
 charge $e$, and $\alpha_{em}$ is the fine-structure constant.

Due to the conservation of the electromagnetic current we can take for real
photons
\be
\sum_r\varepsilon_r^\mu(q)(\varepsilon^{\nu}_r(q))^*\equiv-g^{\mu\nu}
\ee
and thus rewrite the total width as
\begin{eqnarray}
\Gamma&=&\frac{1}{2M}\int
\frac{d^{\,3}P'}{(2\pi)^3\,2E'}
\int \frac{d^{\,3}q}{(2\pi)^3\,2\omega}
\ (2\pi)^4\delta^{(4)}(P-P'-q)
\nonumber\\&&\hspace{2cm}\times
\frac{-1}{2 J+1}\sum_s\ \sum_{s'}
{\cal J}^{BB'\ \mu}_{s\,s'}(P,P')
\left({\cal J}^{BB'}_{s\,s'\ \mu}(P,P')\right)^*
\label{eq:anchura}
\end{eqnarray}
The double sum in Eq.~(\ref{eq:anchura}) is a Lorentz scalar and  it can
only depend on $P^2=M^2,\, P'^2=M'^2$ and  $P\cdot P'=ME'$, with $E'=
(M^2+ {M'}^2)/2M$. As a result, all integrals
can  be done explicitly and we have the final expression
\begin{eqnarray}
\Gamma=\frac{1}{8\pi M^2}\ \frac{M^2-M'^2}{2M}\ \left.\frac{-1}{2 J+1}\sum_s\ 
\sum_{s'}
{\cal J}^{BB'\ \mu}_{s\,s'}(P,P')
\left({\cal J}^{BB'}_{s\,s'\ \mu}(P,P')\right)^*
\right|_{|\vec q\,|=\frac{M^2-M'^2}{2M}}
\label{eq:gammafin}
\end{eqnarray}
where, for the purpose of evaluation, we shall take $\vec q$ along the positive
$Z-$axis.

%
%
%
%
%
%
%
\subsection{Form factor decomposition of the electromagnetic current matrix
elements}
We will  analyze $1/2\to 1/2$ and $3/2\to 1/2$ transitions.

\subsubsection{Case $1/2\to 1/2$}
For $1/2\to 1/2$ transitions we can write the  following form factor
decomposition
\bea
{\cal J}^{BB'\ \mu}_{s\,s'}(P,P')&=&\left\langle B',\ s'\,\vec{P}'=-\vec q\,\left|
J_{em}^\mu(0)\,\right|
\,B,\ s\,\vec P=\vec 0\,\right\rangle\nonumber\\
&=&\bar{u}_{s'}'(-\vec q\,)\left[
\bigg(\gamma^\mu-\frac{2(M-M')P'^\mu}{M^2-M'^2-q^2}\bigg)F_1
+\bigg(
\frac{P^\mu}{M}-\frac{(M^2-M'^2+q^2)P'^\mu}
{M(M^2-M'^2-q^2)}\bigg)F_2\right]\,u_s(\vec 0)
\eea
where $u(\vec 0),\,\bar u'(-\vec q\,)$ are the Dirac spinors
(normalized to twice the fermion mass) for the initial and final
baryon and $F_1,\,F_2$ are form factors that could only depend on the
baryon masses and $q^2$. The above form factor decomposition trivially
satisfies $  q^\mu{\cal J}^{B B'}_\mu = 0$. For the present case we
need the value of the form factors at $q^2=0$ ( $|\vec
q\,|=\frac{M^2-M'^2}{2M}$).

We shall have for the double sum in Eq.~(\ref{eq:gammafin})
\bea
&&\hspace{-2cm}\left.\frac{-1}{2J+1}\sum_s\ \sum_{s'}
{\cal J}^{BB'\ \mu}_{s\,s'}(P,P')
\left({\cal J}^{BB'}_{s\,s'\ \mu}(P,P')\right)^*
\right|_{|\vec q\,|=\frac{M^2-M'^2}{2M}}\nonumber\\
&&=-\frac12Tr\left\{
(\not\!P'+M')\left(
\bigg(\gamma^\mu-\frac{2P'^\mu}{M+M'}\bigg)F_1
+\frac{q^\mu}{M}F_2\right)\right.\nonumber\\
&&\hspace{1.65cm}\left.\left.(\not\!P+M)\ \,
\left(
\bigg(\gamma_\mu-\frac{2P'_{\mu}}{M+M'}\bigg)F_1
+\frac{q_\mu}{M}F_2\right)\right\}\right|_{|\vec
q\,|=\frac{M^2-M'^2}{2M}}\nonumber\\
&&=-\frac12Tr\left\{
(\not\!P'+M')
F_1\bigg(\gamma^\mu-\frac{2P'^\mu}{M+M'}\bigg)
\right.\nonumber\\
&&\hspace{1.65cm}\left.(\not\!P+M)\ \,
\left.
F_1\bigg(\gamma_\mu-\frac{2P'_{\mu}}{M+M'}\bigg)
\right\}\right|_{|\vec q\,|=\frac{M^2-M'^2}{2M}}
\nonumber\\
&&=2\,(M-M')^2\,\left.F_1^2\right|_{|\vec q\,|=\frac{M^2-M'^2}{2M}}\label{eq:traza1}
\eea
where in the second equality we have used current conservation.

The  $F_1$   can be obtained as 
\bea
F_1&=&-\frac{1}{|\vec q\,|\,}\sqrt\frac{E'+M'}{2M}
\,{\cal J}^{BB'\ 1}_{-1/2\,1/2}(P,P')\eea
where we have taken $\vec q$ along the positive $Z-$axis.

\subsubsection{Case $3/2\to 1/2$}
For this case we could use
\bea
{\cal J}^{BB'\ \mu}_{s\,s'}(P,P')&=&\left\langle B',\ s'\,\vec{P}'=-\vec q\,\left|J_{em}^\mu(0)\,\right|
\,B,\ s\,\vec P=\vec 0\,\right\rangle=\bar{u}'_{ s'}(-\vec q\,)
\widehat\Gamma^{\alpha\mu}\,u_{\alpha\ s}(\vec 0)
\eea
where $u_{\alpha}(\vec 0)$ is a Rarita-Schwinger spinor for the
initial spin 3/2 baryon and $\widehat\Gamma^{\alpha\mu}$ is given by
\bea \widehat\Gamma^{\alpha\mu}=\left(
 -\frac{C_3^V}{M'}\left(g^{\alpha\mu} \not\!{q}-
 q^\alpha\gamma^\mu\right) + \frac{C_4^V}{M'^2} \left(g^{\alpha\mu}\,
 q\cdot P- q^\alpha P^\mu\right) + \frac{C_5^V}{M'^2}
 \left(g^{\alpha\mu}\, q\cdot P'- q^\alpha
 P'^\mu\right)\right)\gamma_5 \eea $C_3^V,\,C_4^V,\,C_5^V$ are vector
 form factors that, as before, could only depend on baryon masses and
 $q^2$.  For the double sum in Eq.~(\ref{eq:gammafin}) We have in this case 
 
 \bea 
 &&\left.\frac{-1}{2J+1}\sum_s\ \sum_{s'} {\cal
 J}^{BB'\ \mu}_{s\,s'}(P,P') \left({\cal J}^{BB'}_{s\,s'\
 \mu}(P,P')\right)^*\right|_{|\vec q\,|=\frac{M^2-M'^2}{2M}}\nonumber\\
 &&\hspace{3cm}=\left.-\frac14Tr\left\{
 (\not\!P'+M')\widehat\Gamma^{\alpha\mu}(-1)(\not\!P+M)
 G_{\alpha\beta} \, \gamma^0(\widehat\Gamma^\beta_{\
 \mu})^\dagger\,\gamma^0\right\}\right|_{|\vec q\,|=\frac{M^2-M'^2}{2M}}\ \
\label{eq:traza2}
\eea
with
\be
G_{\alpha\beta}=g_{\alpha\beta}-\frac13\gamma_\alpha\gamma_\beta
-\frac23\frac{P_{\alpha}P_{\beta}}{M^2}
+\frac13\frac{P_{\alpha}\gamma_\beta-P_{\beta}\gamma_\alpha}{M}
\ee
Taking $\vec q$ along the positive $Z-$axis, the
$C_3^V,\,C_4^V,\,C_5^V$ form factors can be obtained as
\bea
C_3^V&=&-\frac{M'}{|\vec q\,|}\sqrt\frac{1}{2M(E'+M')}
\left(\frac1{\sqrt2}\,{\cal J}^{BB'\ 1}_{-3/2\ -1/2}(P,P')
+\sqrt\frac32\,{\cal J}^{BB'\ 1}_{-1/2\,1/2}(P,P')
\right)\nonumber\\
C_4^V&=&\frac{M'^2}{M|\vec q\,|^3}\sqrt\frac{E'+M'}{2M}
\left(\ \ \sqrt\frac32\frac{ME'-M'^2}{M-E'}
\,{\cal J}^{BB'\ 3}_{1/2\,1/2}(P,P')
+\frac1{\sqrt2}\,(2E'-M')\,{\cal J}^{BB'\
1}_{-3/2\ -1/2}(P,P')\right.\nonumber\\
&&\hspace{3.6cm}\left.-\sqrt\frac32M'\,{\cal J}^{BB'\ 1}_{-1/2\,1/2}
(P,P')
\right)\nonumber\\
C_5^V&=&\frac{M'^2}{|\vec q\,|^3}\sqrt\frac{E'+M'}{2M}
\left(-\sqrt\frac32\,{\cal J}^{BB'\ 3}_{1/2\,1/2}(P,P')
-\frac1{\sqrt2}\,{\cal J}^{BB'\ 1}_{-3/2\,-1/2}(P,P')
+\sqrt\frac32\,{\cal J}^{BB'\ 1}_{-1/2\,1/2}(P,P')
\right)
\eea
Within our model we shall obtain (see next section)
\bea
{\cal J}^{BB'\ 3}_{1/2\,1/2}(P,P')=0\ \ ;\ \ 
{\cal J}^{BB'\ 1}_{-3/2\,-1/2}(P,P')=
\sqrt3\,{\cal J}^{BB'\ 1}_{-1/2\,1/2}(P,P')
\eea
so that
\bea
C_5^V=0\ \ , \ \ 
C_4^V=-C_3^V\frac{M'}{M}\ \ ,\ \ C_3^V=-\sqrt\frac32\,\frac{1}{|\vec q\,|}
\sqrt\frac{2M'^2}{M(E'+M')}
\,{\cal J}^{BB'\ 1}_{-1/2\,1/2}(P,P')
\eea
For that case, the trace in Eq.~(\ref{eq:traza2}) can be evaluated  to be
\be
\frac{ (M-M')^2\,(M+M')^4}
{6M^2M'^2}\left.(C_3^V)^2\ \right|_{|\vec q\,|=\frac{M^2-M'^2}{2M}}
\ee
including the $-1/4$ factor.

%
%
%
%
%
%
%
\section{Nonrelativistic states and matrix elements evaluation}
\label{sec:model}
In this section we briefly describe our nonrelativistic states and the
calculation of the electromagnetic current matrix elements within our
model.
\subsection{Nonrelativistic states}
 Our  nonrelativistic states are constructed as 
\begin{eqnarray}
\label{wf}
&&\hspace{-1cm}\left|{B,s\,\vec{P}}\,\right\rangle_{NR}
=\int d^{\,3}Q_1 \int d^{\,3}Q_2\ \sum_{\alpha_1,\alpha_2,\alpha_3}
\hat{\psi}^{(B,s)}_{\alpha_1\,\alpha_2\,\alpha_3}(\,\vec{Q}_1,\vec{Q}_2\,)
\ \frac{1}{(2\pi)^3\ \sqrt{2E_{f_1}2E_{f_2}
2E_{f_3}}}\nonumber\\ 
&&\hspace{3cm}
\times\left|\ \alpha_1\
\vec{p}_1=\frac{m_{f_1}}{\overline{M}}\vec{P}+\vec{Q}_1\ \right\rangle
\left|\ \alpha_2\ \vec{p}_2=\frac{m_{f_2}}{\overline{M}}\vec{P}+\vec{Q}_2\ \right\rangle
\left|\ \alpha_3\ \vec{p}_3=\frac{m_{f_3}}{\overline{M}}\vec{P}-\vec{Q}_1
-\vec{Q}_2\ \right\rangle
 \end{eqnarray}
where $\alpha_j$ represents the spin (s), flavor (f) and color (c)
quantum numbers ( $\alpha\equiv (s,f,c)$\,) of the j-th quark, and
$(E_{f_j},\,\vec{p}_j),\, m_{f_j}$ are its four-momenta and
mass. $\overline{M}$ is given by
$\overline{M}=m_{f_1}+m_{f_2}+m_{f_3}$.

Quark states are normalized such that
\begin{eqnarray}
\left\langle\ \alpha^{\prime}\ \vec{p}^{\ \prime}\,|\,\alpha\
\vec{p}\, \right\rangle=\delta_{\alpha^{\prime}\, \alpha}\, (2\pi)^3\,
2E_f\,\delta^{(3)}( \vec{p}^{\ \prime}-\vec{p}\,)
\end{eqnarray}
In the transitions under study the baryons involved have $b\,c\,l$
quark content, where $l$ represents a light quark $u,d,s$. We choose
the wave functions such that quark 1 is a $b$, quark $2$ is a $c$ and
quark 3 is the light one $l$.
 
 $\hat{\psi}^{\,(B,s)}_{\alpha_1\,\alpha_2\,\alpha_3}
(\,\vec{Q}_1,\vec{Q}_2\,)$ is the internal wave function in momentum
space, being
$\vec{Q}_1$ ($\vec{Q}_2$)  the conjugate momenta to the relative position
$\vec{r}_1$ ($\vec{r}_2$)  between the light quark  and the $b$ ($c$) quark. 
This wave function  is normalized as
\begin{equation}
\int d^{\,3}Q_1 \int d^{\,3}Q_2\ \sum_{\alpha_1,\alpha_2,\alpha_3}
\left(\hat{\psi}^{(B,s')}_{\alpha_1\,\alpha_2\,\alpha_3}(\,\vec{Q}_1,\vec{Q}_2\,)\right)^*
\hat{\psi}^{(B,s)}_{\alpha_1\,\alpha_2\,\alpha_3}(\,\vec{Q}_1,\vec{Q}_2\,)
=\delta_{s'\, s}
\end{equation}
Thus, for our nonrelativistic baryon states we get 
\begin{equation}
{}_{\stackrel{}{\stackrel{}{NR}}}
\left\langle\, {B,s'\,\vec{P}^{\,\prime}}\,|\,{B,s
\,\vec{P}}\,\right\rangle_{NR}
=\delta_{s'\,s}\,(2\pi)^3\,\delta^{(3)}(\vec{P}^{\,\prime}-\vec{P}\,)
\end{equation}
For unmixed states with a well defined $S_h$ value the
$\hat{\psi}^{\,(B,s)}_{\alpha_1\,\alpha_2\,\alpha_3}
(\,\vec{Q}_1,\vec{Q}_2\,)$ wave function has the general form 
\bea
\hspace{-.5cm}\hat{\psi}^{\,(B,s)}_{\alpha_1\,\alpha_2\,\alpha_3}
(\,\vec{Q}_1,\vec{Q}_2\,)=&&\hspace{-.25cm}\frac{\varepsilon_{c_1\,c_2\,c_3}}
{\sqrt{3}!}
\ \widetilde{\phi}^B(\,\vec{Q}_1,\vec{Q}_2\,)\ 
\delta_{f_1\,b}\,\delta_{f_2\,c}\, \delta_{f_3\,l}
(1/2,1/2,S_h;s_1,s_2,s_1+s_2)
(S_h,1/2,J;s_1+s_2,s_3,s)
\eea  
where $\frac{\varepsilon_{c_1\,c_2\,c_3}}{\sqrt{3}!}$ is the color
wave function, with $\varepsilon_{c_1\,c_2\,c_3}$ the fully
antisymmetric tensor in three (color) indices, and the
$(j_1,j_2,j;m_1,m_2,m)$ are Clebsch-Gordan coefficients.

Details on the calculation of the orbital wave function in coordinate space 
for each of the unmixed states involved in this study can
be found in Refs.~\cite{albertusdhb,Albertus:2003sx}. 

\subsection{Matrix elements evaluation}
We evaluate the electromagnetic current matrix elements  as
\bea
{\cal J}^{{ B}{ B}'\ \mu}_{s\,s'}(P,P')&=&
\left\langle { B'},\ s'\,\vec{P}'=-\vec q\,\left|J_{em}^\mu(0)\,\right|
\,B,\ s\,\vec P=\vec 0\,\right\rangle\nonumber\\
&\equiv&\sqrt{2M}\sqrt{2E'}\ 
{}_{\stackrel{}{\stackrel{}{NR}}}\left\langle { B'},\ s'\,\vec{P}'
=-\vec q\,\left|J_{em}^\mu(0)\,\right|
\,B,\ s\,\vec P=\vec 0\,\right\rangle_{NR}\nonumber\\
&=&\sqrt{2M}\sqrt{2E'}\
\left.{\cal J}^{{ B}{ B'}\ \mu}_{s\,s'}(\vec q\,)\right|_{NR}
\eea
with 
\bea
&&\hspace{-.5cm}\left.{\cal J}^{{ B} B'\
\mu}_{s\,s'}(\vec q\,)\right|_{NR}=\sum_j c^B_j\sum_k (c^{B'}_k)^*
  \int d^3Q_1\int d^3Q_2\ \tilde\phi^{{ 
B}}_j(\vec Q_1,\vec Q_2)\nonumber\\
&&\hspace{1cm}\Bigg\{\ \sum_{s_1,s_2}(1/2,1/2,S_{h\,j};s_1,s_2,s_1+s_2)
(S_{h\,j},1/2,J;s_1+s_2,s-s_1-s_2,s)\nonumber\\
&&\hspace{1.5cm}(1/2,1/2,S'_{h\,k};s_1+s'-s,s_2,s_1
+s_2+s'-s)(S'_{h\,k},1/2,J';s_1+s_2+s'-s,s-s_1-s_2,s')\nonumber\\
&&\hspace{1.75cm}\Bigg[-\frac{e}{3}\ \left(\tilde\phi^{ 
B'}_k(\vec Q_1-\frac{m_c+m_l}{\overline{M'}}\,\vec q,\vec Q_2
+\frac{m_c}{\overline{M'}}\,\vec q\,)\right)^*
\ \frac{\bar{u}_{b\
 s_1+s'-s}
(\vec{Q}_1-\vec{q}\,)\ \gamma^\mu\ u_{b\, s_1}(\vec{Q}_1)}{\sqrt{2E_b(|\vec{Q}_1-\vec{q}\, |)2E_b(|\vec{Q}^I_1|)}}\ 
\nonumber\\
&&\hspace{2cm}+\frac{2e}{3}\ (-1)^{S_{h\,j}-S'_{h\,k}}\
\left(\tilde\phi^{B'}_k(\vec Q_1
+\frac{m_b}{\overline{M'}}\,\vec q,\vec
Q_2-\frac{m_b+m_l}{\overline{M'}}\,\vec q\,)\right)^*\ 
\frac{\bar{u}_{c\ s_1+s'-s}(\vec{Q}_2-\vec{q}\,)\ \gamma^\mu\ u_{c\,
s_1}
(\vec{Q}_2)}{\sqrt{2E_c(|\vec{Q}_2-\vec{q}\,
|)2E_c(|\vec{Q}_2|)}}\ \Bigg]\nonumber\\
&&\hspace{1.25cm}+e_{l}\,e\ \left(\tilde\phi^{{B'}}_k(\vec Q_1
+\frac{m_b}{\overline{M'}}\,\vec q,\vec
Q_2+\frac{m_c}{\overline{M'}}\,\vec q\,)\right)^*\
\delta_{S_{h\,j}\,S'_{h\,k}}
\nonumber\\
&&\hspace{2.5cm}\sum_m (S_{h\,j},1/2,J;m,s-m,s)
(S_{h\,j},1/2,J';m,s'-m,s')\nonumber\\
&&\hspace{2.5cm}\times\frac{\bar{u}_{l\ s'-m}(-\vec
Q_1-\vec{Q}_2
-\vec{q}\,)\ \gamma^\mu\ u_{l\, s-m}(-\vec Q_1-\vec{Q}_2)}{\sqrt{2E_l(|-\vec Q_1-\vec{Q}_2-
\vec{q}\, |)2E_l(|-\vec Q_1-\vec{Q}_2|)}}\
\Bigg\}
\label{eq:jnr}
\eea
where we have used the one-body approximation. The first two terms are
the contribution from the $b$ and $c$ quarks respectively, whereas the
third term is the contribution from the light quark. $e_l$ is the
charge of the light quark in units of the proton charge $e$. Besides,
we sum (sums on $j,\,k$) over the different contributions to the
physical states and the $c^B_j,\,c^{B'}_k$ factors are the
corresponding admixture coefficients.  For the evaluation of the
matrix elements we take $\vec q$ along the positive $Z-$axis.

In order to be able to evaluate the spin sums explicitly it is useful
to use the following relations obtained assuming $\vec q$ to be along
the positive $Z-$axis
\bea
&&\hspace*{-.5cm}\frac{1}{\sqrt{2E'2E}}\bar u_{s'}(\vec p\,'=\vec p-\vec q\,)
\gamma^0 u_s(\vec p\,)=
\sqrt\frac{(E'+m)(E+m)}{2E'2E}\chi_{s'}^\dagger\left(\
1+\frac{\vec p\,^2-|\vec q\,|p^3}{(E'+m)(E+m)}\right.\nonumber\\
&&\hspace{8.5cm}\left.+i\frac{|\vec
q\,|}{(E'+m)(E+m)}(\vec\sigma\times\vec p\,)^3\right)\chi_s\nonumber\\
&&=\sqrt\frac{(E'+m)(E+m)}{2E'2E}\left[ \
\left(1+\frac{\vec p\,^2-|\vec q\,|p^3}{(E'+m)(E+m)}\right)\,\delta_{s'\,s}
\right.\nonumber\\
&&\hspace{2.75cm}+\frac{|\vec
q\,|}{(E'+m)(E+m)}\left(\ \, (-p^1+ip^2)\,\delta_{s'\,s+1}\right.\left.
+(p^1+ip^2)\,\delta_{s'\,s-1}\right)\bigg]
\eea
where we work in Pauli-Dirac representation and $\chi$ stands for a Pauli
spinor. Similarly for the spatial components one has 
\bea
&&\hspace*{-.5cm}\frac{1}{\sqrt{2E'2E}}\bar u_{s'}(\vec p\,'=\vec p-\vec q\,)
\gamma^j u_s(\vec p\,)=
\sqrt\frac{(E'+m)(E+m)}{2E'2E}\chi_{s'}^\dagger\left(\ \ \,
\,\frac{\vec p^j}{E+m}+\frac{(\vec p-\vec q)^j}{E'+m}\right.\nonumber\\
&&\hspace{6.5cm}\left.+i\frac{E-E'}{(E'+m)(E+m)}(\vec\sigma\times\vec p\,)^j
-i\frac{1}{(E'+m)}(\vec\sigma\times\vec q\,)^j\right)\chi_s\nonumber\\
&&=\sqrt\frac{(E'+m)(E+m)}{2E'2E}\left[\ \ 
\left(\frac{\vec p^j}{E+m}+\frac{(\vec p-\vec q)^j}{E'+m}+i\frac{E-E'}{(E'+m)(E+m)}
(-p^2\delta_{j1}+p^1\delta_{j2})(\delta_{s\,1/2}-
\delta_{s\,-1/2})\right)\,\delta_{s'\,s}\right.\nonumber\\
&&\hspace{3.75cm}+\delta_{j1}\frac{|\vec q\,|(E+m)-(E-E')p^3}{(E'+m)(E+m)}
\left(\delta_{s'\,s-1}-\delta_{s'\,s+1}
\right)\nonumber\\
&&\hspace{3.75cm}\left.+i\delta_{j2}\frac{|\vec q\,|(E+m)-(E-E')p^3}{(E'+m)(E+m)}
\left(\delta_{s'\,s+1}+\delta_{s'\,s-1}
\right)\right.\nonumber\\
&&\hspace{3.75cm}\left.+\delta_{j3}
\frac{E-E'}{(E'+m)(E+m)}((-p^1+ip^2)\delta_{s'\,s+1}+(p^1+ip^2)\delta_{s'\,s-1})\right]
\eea
Using the above results in Eq.(\ref{eq:jnr}) it is now easy to see why
for $3/2\to 1/2$ transitions we find ${\cal J}^{BB'\
3}_{1/2\,1/2}(P,P')=0$ as a result of the orthogonality relations of the 
Clebsch-Gordan coefficients. Explicit evaluation of the spin sums also shows
that for $3/2\to 1/2$ transitions ${\cal J}^{BB'\ 1}_{-3/2\,-1/2}(P,P')=
\sqrt3\,{\cal J}^{BB'\ 1}_{-1/2\,1/2}(P,P')$. This result can be obtained 
 realizing that 
\bea
&&\sum_{s_1,s_2}(1/2,1/2,1;s_1,s_2,s_1+s_2)
(1,1/2,3/2;s_1+s_2,s-s_1-s_2,s)\nonumber\\
&&\hspace{.5cm}(1/2,1/2,S'_{h};s_1+s'-s,s_2,s_1
+s_2+s'-s)(S'_{h},1/2,1/2;s_1+s_2+s'-s,s-s_1-s_2,s')\ \delta_{s'\ s+1}\nonumber\\
&&\hspace{2.5cm}=-\frac{1}{\sqrt2}\left\langle
[(1/2\otimes1/2)^{S'_{h}}\otimes1/2]^{1/2}_{s'}
\right|
\sigma^{(1)}_{+1}\left|[(1/2\otimes1/2)^{1}\otimes1/2]^{3/2}_s\right\rangle
\eea
where $\left|[(1/2\otimes1/2)^{S}\otimes1/2]^{J}_s\right\rangle$ is
the total spin state of three spin 1/2 particles coupled to total spin
$J$ and third component $s$, and $\sigma^{(1)}_{+1}$ is the $+1$
component of the spin operator of the first particle. Similarly 
\bea
&&\sum_{m}(1,1/2,3/2;m,s-m,s) (S'_h,1/2,1/2;m,s'-m,s')\
\delta_{1\,S'_h}\delta_{s'\ s+1}\nonumber\\
&&\hspace{.5cm}=-\frac{1}{\sqrt2}\left\langle
[(1/2\otimes1/2)^{S'_{h}}\otimes1/2]^{1/2}_{s'} \right|
\sigma^{(3)}_{+1}\left|[(1/2\otimes1/2)^{1}\otimes1/2]^{3/2}_s\right\rangle
\eea with $\sigma^{(3)}_{+1}$ is the $+1$ component of the spin
operator of the third particle. The Wigner-Eckart theorem now gives
\bea &&\left\langle [(1/2\otimes1/2)^{S'_{h}}\otimes1/2]^{1/2}_{-1/2}
\right|
\sigma^{(j)}_{+1}\left|[(1/2\otimes1/2)^{1}\otimes1/2]^{3/2}_{-3/2}\right\rangle\nonumber\\
&&\hspace{3cm}=\sqrt3\ \left\langle
[(1/2\otimes1/2)^{S'_{h}}\otimes1/2]^{1/2}_{1/2} \right|
\sigma^{(j)}_{+1}\left|[(1/2\otimes1/2)^{1}\otimes1/2]^{3/2}_{-1/2}\right\rangle\eea
%
%
%
%
%
%
%
\section{Results and conclusions}
\label{sec:results}
In Table~\ref{tab:gunmixed} we show the results for  the e.m. decay widths
evaluated with the unmixed states of Table~\ref{tab:dhb}, while
\begin{table}
\begin{tabular}{lc||lc}
&$\Gamma (10^{-8}\,{\rm GeV})$&&$\Gamma (10^{-8}\,{\rm GeV})$\\\hline
$\Xi^*_{bcu}\to\Xi_{bcu}'\gamma$\hspace{.25cm} &  $4.04$ 
&\hspace{.25cm}$\Omega_{bc}^*\to\Omega_{bc}'\gamma$\hspace{.25cm} & $3.69$ \\ 
$\Xi^*_{bcd}\to\Xi_{bcd}'\gamma$\hspace{.25cm} &  $4.04$ 
& &  \\ 
$\Xi_{bcu}^*\to\Xi_{bcu}\gamma$
&$105$
&\hspace{.25cm}$\Omega_{bc}^*\to\Omega_{bc}\gamma$
&$20.9$\\ 
$\Xi_{bcd}^*\to\Xi_{bcd}\gamma$
&$50.5$
&
&\\ 
$\Xi_{bcu}'\to\Xi_{bcu}\gamma$ &  $0.992$ 
&\hspace{.25cm}$\Omega_{bc}'\to\Omega_{bc}\gamma$ & $0.568$ \\ 
$\Xi_{bcd}'\to\Xi_{bcd}\gamma$ &  $0.992$ 
& &  \\ 
\hline
\end{tabular}
\caption{ Electromagnetic decay widths (in units of $10^{-8}\ {\rm
 GeV}$) for unmixed states with  a well defined $S_h$ value.}
 \label{tab:gunmixed}%
\end{table}
in the left panel of Table~\ref{tab:gmixed}, the results using the
physical spin-1/2 $bc$ states of Table~\ref{tab:dhbphys}, are
given. The effects of mixing are relevant for all transitions. In
particular we find 
\bea
\Gamma(\Xi^*_{bcd}\to\Xi_{bcd}^{(1)}\gamma)&\approx&\frac{1}{33}\
\Gamma(\Xi^*_{bcd}\to\Xi_{bcd}'\gamma)\nonumber\\
\Gamma(\Xi_{bcu}^{(1)}\to\Xi_{bcu}^{(2)}\gamma)&\approx& 13\
\Gamma(\Xi_{bcu}'\to\Xi_{bcu}\gamma)\nonumber\\
\Gamma(\Xi_{bcd}^{(1)}\to\Xi_{bcd}^{(2)}\gamma)&\approx& 21\
\Gamma(\Xi_{bcd}'\to\Xi_{bcd}\gamma)\nonumber\\
\Gamma(\Omega^*_{bcd}\to\Omega_{bcd}^{(1)}\gamma)&\approx&\frac{1}{12}\
\Gamma(\Omega^*_{bcd}\to\Omega_{bcd}'\gamma)\nonumber\\
\Gamma(\Omega_{bc}^{(1)}\to\Omega_{bc}^{(2)}\gamma)&\approx& 15\
\Gamma(\Omega_{bc}'\to\Omega_{bc}\gamma) 
\eea 
which shows very clearly
that hyperfine mixing can not be ignored when evaluating
e.m. transitions involving spin-1/2 $bc$ baryons. Besides, we observe
that mixing breaks the degeneracy originally present in the unmixed
case for most transitions involving $bcu$ and $bcd$ baryons.
\begin{table}
\begin{tabular}{lc||lc}&$\Gamma (10^{-8}\,{\rm GeV})$&&$\Gamma (10^{-8}\,{\rm GeV})$\\\hline
$\Xi^*_{bcu}\to\Xi_{bcu}^{(1)}\gamma$ &  $6.05$ 
&\hspace{.25cm}$\Omega_{bc}^*\to\Omega_{bc}^{(1)}\gamma$ & $0.31$ \\ 
$\Xi^*_{bcd}\to\Xi_{bcd}^{(1)}\gamma$ &  $0.12$ 
& &  \\ 
$\Xi_{bcu}^*\to\Xi_{bcu}^{(2)}\gamma$
&$73.9$
&\hspace{.25cm}$\Omega_{bc}^*\to\Omega_{bc}^{(2)}\gamma$
&$50.2$\\ 
$\Xi_{bcd}^*\to\Xi_{bcd}^{(2)}\gamma$
&$103$
&
&\\ 
$\Xi_{bcu}^{(1)}\to\Xi_{bcu}^{(2)}\gamma$ &  $12.4$ 
&\hspace{.25cm}$\Omega_{bc}^{(1)}\to\Omega_{bc}^{(2)}\gamma$ & $8.52$ \\ 
$\Xi_{bcd}^{(1)}\to\Xi_{bcd}^{(2)}\gamma$ &  $20.9$ 
& &  \\ 
\hline
\end{tabular}\hspace{.5cm}
\begin{tabular}{lc||lc}&$\Gamma (10^{-8}\,{\rm GeV})$&&$\Gamma (10^{-8}\,{\rm GeV})$\\\hline
$\Xi^*_{bcu}\to\Xi_{bcu}^{(1)}\gamma$ &  $1.56$ 
&\hspace{.25cm}$\Omega_{bc}^*\to\Omega_{bc}^{(1)}\gamma$ & $0.415$ \\ 
$\Xi^*_{bcd}\to\Xi_{bcd}^{(1)}\gamma$ &  $0.748$ 
& &  \\ 
$\Xi_{bcu}^*\to\Xi_{bcu}^{(2)}\gamma$
&$123$
&\hspace{.25cm}$\Omega_{bc}^*\to\Omega_{bc}^{(2)}\gamma$
&$24.2$\\ 
$\Xi_{bcd}^*\to\Xi_{bcd}^{(2)}\gamma$
&$59.2$
&
&\\ 
$\Xi_{bcu}^{(1)}\to\Xi_{bcu}^{(2)}\gamma$ &  $22.8$ 
&\hspace{.25cm}$\Omega_{bc}^{(1)}\to\Omega_{bc}^{(2)}\gamma$ & $3.78$ \\ 
$\Xi_{bcd}^{(1)}\to\Xi_{bcd}^{(2)}\gamma$ &  $11.0$ 
& &  \\ 
\hline
\end{tabular}
\caption{ Electromagnetic decay widths (in units of $10^{-8}\ {\rm
 GeV}$) for physical states. We show the full calculation results
 (left panel) and results obtained considering only the contributions 
 where the total spin of the heavy
 quark subsystem does not change (right panel).}
 \label{tab:gmixed}%
\end{table}

We know that in the infinite heavy quark mass limit only the terms
where the spin of the heavy quark subsystem does not change can
contribute to the decay widths. This effect can already be seen in
Table~\ref{tab:gunmixed} in the fact that
$\Gamma(\Xi_{bcl}^*\to\Xi_{bcl}\gamma)\gg
\Gamma(\Xi_{bcl}^*\to\Xi_{bcl}'\gamma)$ or in the smallness of
$\Gamma(\Xi_{bcl}'\to\Xi_{bcl}\gamma)$.  Similar results are observed
in the $\Omega$ sector.
 
To see how far from the ideal infinite heavy quark mass limit we are
in this case, we show in the right panel of Table~\ref{tab:gmixed} the
results obtained with the physical spin-1/2 $bc$ states, but
considering only the contributions to the decay widths of the terms
where the total spin of the heavy quark subsystem does not change. We
see big changes when compared to the full results in the left panel of
the same Table. As a result, and in contrast to the weak decay case
discussed in Ref.~\cite{albertus2010}, and to our prior expectations
also outlined in this latter reference, heavy quark spin symmetry
relations deduced in the infinitely heavy mass limit,  are not accurate enough
for the study
of e.m. transitions involving doubly heavy baryons.
Next-to-leading corrections turn out to be quite large as the differences
between the two panels in Table~\ref{tab:gmixed} indicate. Another important
point here is that the decay widths are proportional to the factor
$(M^2-M'^2)(M-M')^2$ coming from phase space and the spin sums. As $M$
is close to $M'$, the decay widths are very sensitive to the actual
baryon masses.

Finally, we would like to stress, once more, that the experimental
measurement of e.m. widths will be extremely valuable in order to
extract information on the hyperfine mixing of doubly heavy $bc$
baryons, as the difference among the results of
Table~\ref{tab:gunmixed} (for unmixed states) and the left panel of
Table~\ref{tab:gmixed}
(mixed states) clearly show.  However, we should also point out that
by looking only at e.m. transitions, it would
not be possible to determine their actual mixing matrix without
relying on a theoretical model. In this respect, the situation is more
favorable in the case of the semileptonic weak decays of these
baryons, as we discussed in Ref.~\cite{albertus2010}, where leading
order heavy quark symmetry relations turned out to be much more
accurate.

%
%
%
%
%
%
%
\begin{acknowledgments}
  This research was supported by DGI and FEDER funds, under contracts
  FIS2008-01143/FIS, FIS2006-03438, FPA2007-65748, and the Spanish
  Consolider-Ingenio 2010 Programme CPAN (CSD2007-00042), by Junta de
  Castilla y Le\'on under contracts SA016A07 and GR12, by Generalitat
  Valenciana under contract PROMETEO/20090090 and by the EU
  HadronPhysics2 project, grant agreement n. 227431. 
\end{acknowledgments}

\end{document}